%% file: arxiv.tex
\documentclass{article}
\usepackage{arxiv}

\usepackage[utf8]{inputenc} % allow utf-8 input
\usepackage[T1]{fontenc}    % use 8-bit T1 fonts
\usepackage{hyperref}       % hyperlinks
\usepackage{url}            % simple URL typesetting
\usepackage{booktabs}       % professional-quality tables
\usepackage{amsfonts}       % blackboard math symbols
\usepackage{nicefrac}       % compact symbols for 1/2, etc.
\usepackage{microtype}      % microtypography
\usepackage{lipsum}
\usepackage{cite}
\usepackage{amsmath,amssymb,amsfonts}
\usepackage{algorithmic}
\usepackage{graphicx}
\usepackage{textcomp}
\usepackage{todonotes}
\usepackage{xcolor}
\usepackage{graphicx}
\graphicspath{ {./images/} }

\title{Cyberphysical Sequencing for Distributed Asset Management with Broad Traceability}

\author{
 Joshua E. Siegel$^1$ \\
  Department of Computer Science and Engineering\\
  Michigan State University\\
  East Lansing, MI 48824 \\
  \texttt{jsiegel@msu.edu} \\
   \And
 Gregory Falco$^1$ \\
  Institute for Assured Autonomy (IAA)\\
  Johns Hopkins University\\
  Baltimore, MD 21211 USA\\
  \texttt{falco@jhu.edu} \\
}

\begin{document}
\maketitle
\footnotetext[1]{Both authors contributed equally to this manuscript.}

\begin{abstract}
Cyber-Physical systems (CPS) have complex lifecycles involving multiple stakeholders, and the transparency of both hardware and software components' supply chain is opaque at best. This raises concerns for stakeholders who may not trust that what they receive is what was requested. There is an opportunity to build a cyberphysical titling process offering universal traceability and the ability to differentiate systems based on provenance. Today, RFID tags and barcodes address some of these needs, though they are easily manipulated due to non-linkage with an object or system's intrinsic characteristics. We propose cyberphysical sequencing as a low-cost, light-weight and pervasive means of adding track-and-trace capabilities to any asset that ties a system's physical identity to a unique and invariant digital identifier. CPS sequencing offers benefits similar Digital Twins' for identifying and managing the provenance and identity of an asset throughout its life with far fewer computational and other resources. 
\end{abstract}

% keywords can be removed

\input{content}

\end{document}

%% file: content.tex
\section{Why Pervasive Digital Traceability (PDT)?}
\input{opportunity}

\section{Elements of Enhanced Asset Management}\label{explor}
\input{elements}

\section{What PDT Might Look Like}\label{whatitdo}
\input{proposed}

\section{Opportunities, Challenges, and Limitations}
\input{challenges}

\section{Conclusion}
\input{conclusion}

\bibliography{references}
\bibliographystyle{IEEEtran}

%% file: opportunity.tex
Across domains, manufactured and assembled system complexity is increasing. Constituent components require compliance with stringent specifications, must have low defect rates, and increasingly require known provenance relating to origin and interaction histories. At the same time, economic and other constraints affecting production and assembly may necessitate involving diverse and untrusted vendors: a vehicle's parts may be made abroad and assembled domestically, while a medication might be compounded in one country before being shipped to another for packaging and a third for distribution. Power generation plant components might be manufactured globally but require certification in the country of use, while electronics manufacturing for a globally-distributed device may require trust-related integrated circuits to be provided and validated by a single-source vendor. 

Diverse and distributed supply chains invite substantial risk of counterfeit, compromise, or non-compliance. Providing insight into component origin, provisioning and life before assembly has the potential to make resulting systems more robust, resilient, and broadly usable than is feasible today. Consistent validation of the trusted identity and integrity of a system is necessary to support system tracking, supervision, management, and transferrability to mitigate this risk, particularly for critical applications. 

Asset supervision may be simple: high assurance systems like space vehicles have robust, metadata-abundant supply chains. In other cases, system, sub-assembly, and component mapping is more difficult. Assembled automobiles, for example, belong to a global registry for Vehicle Identification Numbers that supports asset supervision and management – but outside of engines and structural elements, components themselves are may not be universally serialized or tracked after sale. The smaller and lower cost a component, the less likely it is to be uniquely identifiable. This poses challenges particularly as low-cost electronics increasingly support driver assistance and other safety-critical technologies. The issue of non-unique identification and non-traceability is worsened for smaller, lower-cost assets such as consumer electronics and appliances, where the cost of serialization and immutable identification is high relative to the cost of the component, assembly, or system. 

Irrespective of the instrumented system or component, broad asset identification, traceability, supervision, and management offer the potential for substantial safety, economic, reliability, and usability benefits in addition to providing information necessary for supply chain optimization. These capabilities are best taken advantage of in a \textit{digital} context; digital systems offer heightened transparency and ease of data access and sharing relative to physical-only approaches, with the potential to scale substantially. For example, tracing individual medical pills electronically would allow for manufacturers, doctors, and patients to understand the origin and chain of custody of a pill and regularly enforce compliance with medical prescriptions, whereas tracing a bottle of pills on a written ledger only allows a pharmacist to verify that the pills are ostensibly of sound origin and have or have not been distributed to the appropriate individual at the right time. 

In view of the unmet need to instrument smaller and lower cost assets digitally, there are safety, security, economic, and social advantages in turning every physical system into uniquely differentiated Cyber-Physical System (CPS) comprehensively, efficiently, and with minimal manual intervention. System classification, characterization, and identification techniques built upon mechanical and software engineering capabilities have the potential to extract differentiating features for diverse systems, thereby facilitating automated system identification. Pairing unique identifiers with an instance-specific, immutable digital identifier associated with lightweight asset mirrors can enable automated non-repudiable track-and-trace capabilities supporting critical applications including smart supply chain management, digitalized and networked logistics systems, AI-backed logistics decision support systems, digital forensics, and safety-critical Healthcare and Industry 4.0 applications. 

In this article, we detail one potential embodiment of a universal, automated asset digitization system suitable for providing enhanced supply chain, component provenance, and system state data and explore how these capabilities will contribute to a future in which universal, low-cost and low-touch asset mirroring can enable the benefits of pervasive supply chain unrealizable with contemporary technology. 

We begin by identifying the opportunity for pervasive universal asset identification, and then explore technical elements necessary to enable such enhanced asset management. We then present representative components and architecture for a lightweight system for asset tracking combining these affordances in a holistically-optimized manner, and close with a discussion of challenges, opportunities, and future directions for creating a system capable of universally mirroring physical systems in a digital context.

%% file: elements.tex
Traditional supply chain management techniques require frequent manual intervention or operate on systems and assets of limited diversity, challenging their ability to massively scale. Human intervention takes the form of manual counts or supervised automated scanning, data entry, model creation, and digital system supervision and management. Creating a sustaining and scaleable asset management solution necessitates the development and deployment of novel automated system identification techniques, resource-light pervasive digital mirroring capabilities, and secure and effective large-scale mirror management techniques. We detail the need for and role of each such capability in this section.

\subsection{System Identification}\label{sysid}
Highly-specific system identification is essential to the tracking and traceability of individual system instances without manual intervention. Automated identification, which uses software to differentiate among system types and instances, relies primarily on two algorithmic processes: classification, which identifies the system type, and characterization, which provides quantitative and qualitative measures of a specific system's state. Classification narrows a computer's need to search a broad range of assets to identify a system; this of this like playing ``20 Questions'' - few attributes greatly narrow the number of possible systems. Once the search area is focused, precise system characterization may be used to uniquely identify an instance of a particular system class; for example, once a computer knows that an asset to be identified is a car, it can look at the license plate to resolve the car to a particular instance. Another example using joint classification and characterization for unique identification is facial recognition: first, a human is identified, then, a face, and then facial landmarks are used to identify a specific person. In the case of identifying an appliance, an algorithm might first determine that the sound it hears must come from a machine with a motor, then a washing machine, then a washing machine on the spin cycle - and from subtle noises reflective of the belt's wear and a slight imbalance in the manufactured drum, know where that particular machine is located and its serial number. 

Classification and characterization may use input data from proprioceptive (self-measuring) or exterioceptive (externally-measuring) sensors such as those found in smart phones~\cite{KANARACHOS2018867}. For example, a smartphone microphone might be used as an extereoceptive sensor to identify a particular car engine~\cite{siegel_sae_vibro} that can be mapped to unique Vehicle Identification Number. Or, a car may use proprioceptive accelerometers to identify the serial number of a tire that is part of an imbalanced wheel and tire assembly~\cite{wheel_imbalance}. A bandage might be identifiable to an exterioceptive sensor such as a laptop's webcam using a technique such as DigiMark~\cite{digimark}, or with sufficient resolution, through ``witness marks'' that serve as visual fiducal indicators left behind unintentionally as a manufacturing artifact. 

Once captured, sensor data are processed, often using techniques such as \textit{frequency analysis}, \textit{cepstrum analysis}, \textit{filtering}, \textit{wavelet analysis}, and other approaches such as computer vision. These techniques generate robust features that are resistant to small perturbations such that algorithms can easily and reliably identify unique landmarks without model overfit, even as the system's state evolves. These landmarks and features may be extracted without prior knowledge or human intervention~\cite{SHEN2018170}.

The algorithmic foundations for classification and characterization vary by domain and are well-established in literature~\cite{siegel_misfire,siegel_air_filter,wheel_imbalance,siegel_tire_pressure,siegel_tire_cracks,siegel_ms_thesis_maybe,siegel_sae_vibro}. Importantly, classification and characterization algorithms are transferrable across systems, contexts~\cite{siegel_ms_thesis_maybe,7328363,5482295,8595428,8793032,10.1007/978-3-030-29513-4_11,gozick2012driver,7130669,sharma2019pothole,eriksson2008pothole,chugh2014road,aleadelat2018estimation,kyriakou2021vehicles,10.1007/978-3-319-91635-4_7,abdic2016detecting,ZHAO201992,sedivy2019mechatronics,8735446,cherian2019mobile,vij2018smartphone,10.1007/978-3-319-61461-8_4,9340328,chiou2019survey,sawczuk2016application,sun2019strategy,iannace2019fault,barlow2019using,4273732} and domains~\cite{glowacz2019,10.1007/978-3-540-24646-6_2,1677514,4487086,10.1145/2370216.2370269,kwapisz2011activity,doi:10.1155/2014/503291,6343802,ellis2015validated}.

Detection of micro-scale faults and imperfections is suitable for unique instance identification~\cite{siegel2020surveying} for some asset classes. With broad model coverage, automated system identification may therefore be used to uniquely serialize diverse physical systems without human intervention. This would eliminate the substantial human bottleneck of asset identification, serialization, and matching with digital representations, thereby supporting the tracking and traceability of larger-scale, more diverse systems than presently possible. 

\subsection{Pervasive Mirroring and System Digitization}\label{permir}
Interconnected CPS present opportunities for large-scale cross-context data collection, analytics, and actuation that augment, extend, and complement the supply chain advances enabled by pervasive digital identification. As a result, CPS increasingly represent physical systems mirrored within remote computing endpoints. 

One way of mirroring system parameters is using a Digital Twin (DT). A DT is a virtual entity connected to a real world entity comprising three elements: the \textit{twin} itself comprises the digital ``identity'' and instance of a system, the \textit{connection} links the digital with the physical, and then a \textit{model} mirrors the physical entity digitally using reflective state-space and metadata representations. DT's can be diversely characterized~\cite{AutiosaloVepsalainenViitalaEtAl2020, 9108291, JONES202036, 8901113, QI20213, 10.1109/SEsCPS.2019.00012, 7819217, van2020taxonomy} but in general aim to richly mirror a system \textendash capturing useful data at the expense of significant computational, energetic, and network resources. Often, DT's and physical systems are matched and modelled manually, the effort of which limits their scale. Reducing resource requirements and automating model creation and matching has the potential to enhance digital system mirrors' utility in supply chain and other applications by providing enhanced full-chain insight into system state, performance, and context among larger networks. 

Data Proxies~\cite{siegel2016data,siegel2018cognitive,siegel2019cognitive,siegel2020cloud} offer an alternative solution to Digital Twins that address resource and modelling challenges facing Twins' large-scale adoption. Like Twins, Proxies are rich representations of physical or other systems. Unlike Twins, Proxies use system models to recreate faithful representations from sparse - and therefore, resource-light - data. Compared with DT's, Proxies are more resource efficient and their execution within remote computing environments supports enhanced and scaleable context-aware security~\cite{siegel2016data,siegel2018cognitive,siegel2019cognitive,siegel2020cloud} capable of protecting both data streams and diverse, connected systems. In economic terms, while Proxies seek a ``pareto solution'' to Digital Twinning that attains a high-fidelity mirror with significantly reduced resource requirements. 

Proxies may also be augmented with additional data types: supply chain and logistics management, for example, requires understanding not only the what, where, and how of a system at present time, but also system and component provenance. Provenance may inform judgements on trustworthiness and accuracy of a system, based on a history of data acquisition, modification, maintenance, and custody (but not ownership). Data Proxies can benefit substantially from the capture and service of metadata. 

Data provenance is a well-established concept in some domains, but is lesser-explored in CPS. Today, provenance is difficult to share efficiently - and must be thoughtfully designed and digitally represented so as not to tax a network. At the same time, the provenance mechanism must be secured to assure full-chain trustworthiness~\cite{reznik2013poster,sultana2012comprehensive} A simple provenance model encapsulates the source of data, their capture and transmission mechanisms - and intermediate handlers - and modifications occurring at points between the sensor and the end-user or end-use application.  

There have been efforts to standardize provenance reporting for data in connected devices, e.g. ECMA TR/110\cite{tc53pscp} which defines reportable metadata as including peripheral sensor details, identification of the physical variant of a sensing device, information derived from the clock, and geospatial data. In the context of logistics and supply chain, additional parameters may be mirrored. 

If systems are uniquely and automatically identified as described in Section~\ref{sysid}, classification results may be used to create and pair systems with appropriate mirroring models for recreating and representing the system's state and metadata using few computational resources. In this manner, low-touch physical system mirroring will support automated, detailed asset tracking and supply chain management as well as broader distributed system applications and optimizations. 

\subsection{Mirror Management}
Digital Twins are often mirrored within proprietary databases using locked-down Application Programming Interfaces (APIs) to facilitate access. These Twins mirror certain system types and enable particular applications and use cases. In comparison, Distributed Ledger Technology (DLT) offers advantages related to scalability, ease of access, and security. DLT is a class of decentralized multi-party systems that operate synchronously. There are several types of DLT, notably blockchain. Blockchain is a data structure that consists of blocks of data linked in a chain by cryptographic hashes. 

DLT enables a secure means for consensus-driven recording, even in adversarial environments where malicious actors attempt to change values or disrupt the ledger. Data may include asset identifiers, titles, metadata, or most any digital information in readily-accessible or encrypted form. To successfully alter the ledger, a malicious actor must control $51$\% of processing power comprising the ledger's mining infrastructure; large-scale attacks are unlikely.

Blockchain's uses have grown since Bitcoin, a decentralized payment system, and Ethereum, a means for establishing smart multi-party contracts, were introduced. Ethereum's smart contract platform allows software developers to build applications such as title tracking or smart contracts. Other applications secure and validate software updates\cite{steger2018secure,dorri2017blockchain,Chanson:2017:BPE:3123024.3123078,10.1007/978-3-319-66972-4_12}, log readings and system identities\cite{brousmiche2018digitizing,chanson2017blockchain, 11-03-02-0006}, manage credentials\cite{malik2018blockchain} and data, and provide authentication\cite{moinet2017blockchain}. Blockchain has also been used to store and secure lifecycle data, service records, and accident histories and reconstruction from supply chain through end-of-life\cite{brousmiche2018digitizing,anwar2019ensuring,8493118,8626103}.

A recently-popular DLT application is Non-Fungible Tokens (NFTs)~\cite{entriken2018eip}. Unlike other assets, such as Bitcoin, NFTs are non-fungible: the digital identifier is linked inextricably to a particular instance of a ``thing.'' DLT and NFTs therefore have the potential to assign immutable, non-revocable identifiers to specific digital and physical things, allowing the creation of a mass-scale set of identifiers for diverse systems. These identifiers can be linked not only to the identity of a physical thing, but also to all other attributes of a Digital Twin or Data Proxy -- connecting the system mirror itself to a pervasive and secure universal identifier accessible to anyone with appropriate credentials. 

With the thoughtful combination of DLT and NFTs with Digital Twins, Data Proxies, and/or provenance metadata (as described in Section~\ref{permir}, unique digital identifiers may be tied inextricably to specific physical and other system instances -- creating a faithful, secure, and broadly-accessible mirror of the real world in a digital context. 

Individually, these technologies address challenges and unmet opportunities in enabling digital supply chain and logistics; combined thoughtfully, these technologies will underpin a system supporting pervasive and universal-scale system mirroring and supervision that will lead massive societal benefit. We describe a vision for a holistically-developed set of these contituent technologies in Section~\ref{whatitdo}.

%% file: proposed.tex
As Section~\ref{explor}'s exploration of the elements of enhanced asset management demonstrates, there have been substantial efforts in the areas of automatically identifying, efficiently mirroring, and securely and scalably storing records for physical assets - but such elements have not been combined for the mirroring of diverse systems, at scale, with high granularity and minimal human involvement. The holistic development of a unified and automated digital asset management solution could enable novel supply chain and logistics optimization capabilities, as well as support the development of data-enhanced management and supervision applications. In this section, we present a vision for a holistically-designed solution merging these enabling elements in support of the complete and comprehensive mirroring of all manner of physical and other systems and their constituent components. 

For some use cases, elements of this holistic approach exist - barcodes or RFID tags encode information that may uniquely identify a system to enable track-and-trace capabilities, while a Vehicle Identification Number may encode information about the origin of that vehicle (country of manufacture, options, etc.). However, these existing techniques do not rely on system intrinsic metrics to establish identity \textendash a label may be separated from the system, presenting questions of security and trust. Further, tagging and data entry are often manual or otherwise limited to systems that have low-variability to facilitate automatic serialization. No extant approaches combine unique, invariant identification with observable and low-mutability origin and provenance markers to enable mass-scale asset tracking and management. 

In presenting a vision for a system addressing these unmet opportunities, we first draw a parallel to biology: in the case of a bioorganism, the genome determines the composition of the organism as well as provides a unique identifier, while epigenetics codifies the way in which environmental conditions and behaviors impact that genome's expression. In a (cyber)physical system, a ``physical hash'' may offer an analog that invariantly identifies a system and the composition of that system, and its contextual, deterministic ``expression'' based on the operational lifecycle of the system. In this manner, the physical hash can be used to identify - throughout its lifecycle - particular instances of a system or component. 

Just as genomes can be sequenced to provide identifying information and insight into biological systems, we propose a ``CPS Genome Project'' for automatically matching a physical system with a digital representation of that system and instance, enabling security and identity verification and facilitating trusted CPS transactions. Rather than looking at DNA, this Project will evaluate observable physical hashes to identify and evaluate particular systems and comprise a system-specific digital identifier inextricably linked to that specific asset. Features unique to the system may be captured by on-system or ambient sensors. This will enable any physical system, with or without sensing, connectivity, or networking, to become cyberphysical in nature without human involvement. This will facilitate system identification, tracking, supervision, management, and transferrability. An instance-specific ``metadata bundle'' will describe the system or component's particular provenance and lifecycle. The system itself, and its metadata, may be mirrored as a Data Proxy referred to by an immutable DHT for asset management. 

Such a system would draw upon elements of proprioceptively- and exterioceptively-sensed system classification, characterization, and identification\cite{siegel_misfire,siegel_air_filter,wheel_imbalance,siegel_tire_pressure,siegel_tire_cracks,siegel_ms_thesis_maybe}, a hybridized distributed ledger architecture for communicating the operations of a complex system\cite{11-03-02-0006}, and metadata tagging\cite{tc53pscp}. The thoughtful combination of such techniques with clean-sheet design will support the fast and effective development of CPS sequencing, metadata bundling, and a DLT/NFT technology for mirroring identities and transaction histories robustly at scale.

Classifying a system uses diverse sensors to answer a series of questions, with each subsequent question narrowing the search space. The below excerpt is from \url{http://www.20q.net}, when the example subject asset to be tracked is a physical key:
\noindent\fbox{%
    \parbox{\linewidth}{%
\textit{Is it Animal, Vegetable, Mineral, Concept, or Unknown?} Unknown \\
\textit{Is it very large?} No \\ 
\textit{Do you use it at work?} Yes\\
\textit{Is it black?} Rarely\\
\textit{Can it be used to write?} Doubtful\\
\textit{Is it hard?} Yes\\
\textit{Does it like to run?} No\\
\textit{Can liquids pass through it?} No\\
\textit{Would you use it in the dark?} Sometimes\\
\textit{Is it very, very expensive?} No\\
\textit{Is it round?} No\\
\textit{Does it usually have four corners?} No\\
\textit{Can it be placed on your head?} Yes\\
\textit{Does it have a hole in it?} Usually\\
\textit{Does it break if dropped?} No\\
\textit{Is it flexible?} No\\
\textit{Does it have writing on it?} Sometimes\\
\textbf{\textit{It's probably a key}}
    }%
}

\begin{figure}
    \centering
    \includegraphics[width=0.4\linewidth]{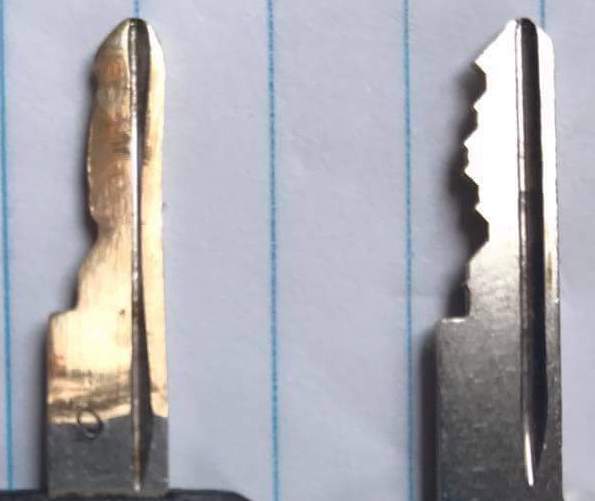}
    \caption{The left key and right key have the same cut pattern; the left key is considerably more worn-down.}
    \label{keys}
\end{figure}

From 20 questions that can largely be answered from low-cost sensors, the website correctly identified the class of system in question. In the case of the example system embodiment, this information could primarily be observed from a camera external to the key itself - which would ultimately allow the system to be mirrored without any onboard sensing, computation, or connectivity, relying instead of pervasive sensing and computation for instrumentation, analysis, and reporting. Next, the \textit{specific} system needs to be identified. To do this, some class-specific knowledge may be applied - in this case, we know that keys wear with use, and the wear on each cut (``witness marks'') can tell us something about the key's history - and its associated lock. See Fig.~\ref{keys} for examples of how key cuts might wear with use and age.

In the case of two keys that are identical by design but that age differently, such differences can be characterized using the same or different sensors from those that identified the class of the subject asset. Here, we can again use vision to assess the relative age and wear of each key, with potential other insights into provenance such as likely material, and based on other visual artifacts, whether the key in question is original or a copy. With sufficient data, the key can be tied to a particular instance of an asset, even if that asset might have many near-identical copies.

\begin{figure}
    \centering
    \includegraphics[width=0.4\linewidth]{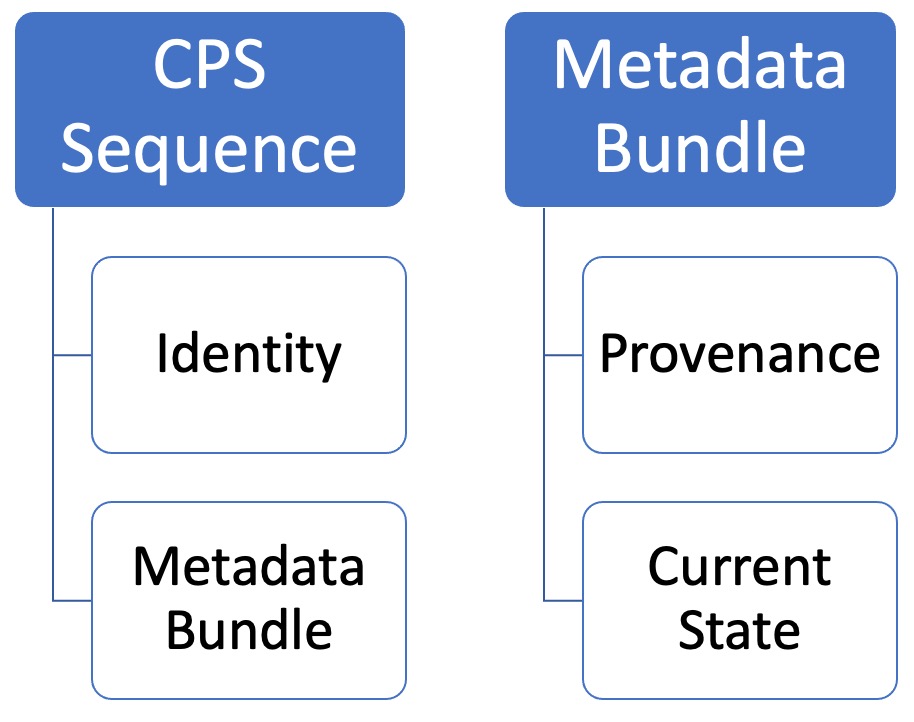}
    \caption{Identity and a metadata comprise the CPS sequence; the metadata bundle itself comprises both implicit and explicit provenance and observed measures for the system's current state.}
    \label{heirarchy}
\end{figure}

Once the key has been uniquely identified, its current state and provenance are combined into a metadata bundle. The identity, state, and provenance then comprise the ``Cyberphysical Sequence'' of that asset. The heirarchical relationship among the CPS Sequence, Identity, the Metadata Bundle, Provenance, and the Current State (all as determined by pervasive sensing) is represented in Figure~\ref{heirarchy}.

\begin{figure}
    \centering
    \includegraphics[width=0.8\linewidth]{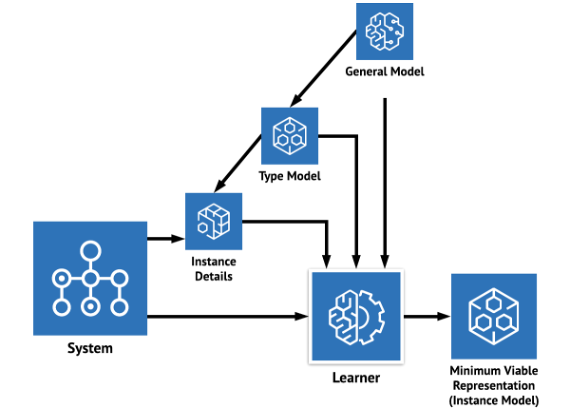}
    \caption{When a system is mirrored, it first uses a generalized model. Over time, models are adapted and leaned to mirror critical parameters with minimal resources.}
    \label{adapt}
\end{figure}

\begin{figure}
    \centering
    \includegraphics[width=0.9\linewidth]{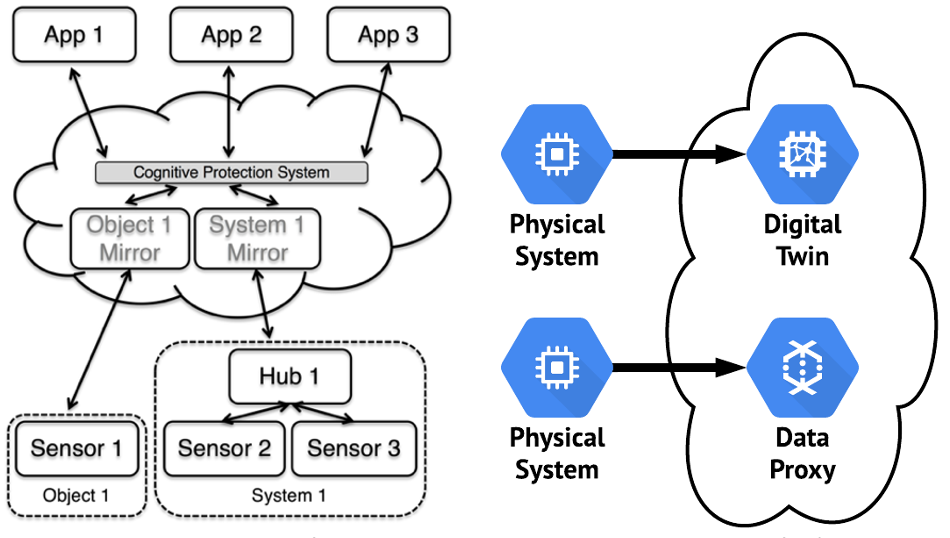}
    \caption{Left: the mirrors (center) represent Data Proxies within a Cloud computing environment. The Proxy architecture allows diverse systems and applications to co-mingle with limited resource expenditure and few areas for interstitial weakness. Right: Proxies store sparse representations of systems in remote computing environments, whereas Digital Twins offer higher resource intensity, more faithful mirrors. }
    \label{proxyarch}
\end{figure}

Once the system has been uniquely identified, that identity may then me associated with a Data Proxy. This Proxy begins with a high-fidelity mirroring approach and observes system and interaction dynamics, adapting to learn an efficient state-space reconstruction model appropriate for the measurable and hidden attributes of the asset, in this case, a key. We call this the ``minimum viable representation,'' and it depends both on the system being mirrored and the end-use applications for its data. The adaptation from general to instance-specific model is shown in Fig.~\ref{adapt}, while the Proxy architecture is shown in Fig.~\ref{proxyarch} and is explored in-depth in \cite{8055694}. In the case of a key, one might imagine that the 

Proxies resulting from CPS sequencing will differ from conventional Digital Twins in that most Twin models are structured based on a particular asset type and tailored through manual data entry and other associative activities, whereas Proxies' originating assets are automatically identified and efficiently mirrored using a generalized, tailorable system model augmented with broader metadata reflecting provenance in addition to the system's current state. These Proxies are dynamically instantiated, rather than explicitly created, and integrated with DHT implementations, may be used to validate trusted identity more robustly than is presently feasible. This allows for the digitization of assets focused equally as much on whether a system ``is'' or ``is not'' that requested as much as faithfully mirroring that system throughout its lifecycle. Computationally-lightweight sequencing will allow all physical systems to have a unique digital identifier, irrespective of computational, sensory, or network constraints. 

\begin{figure}
    \centering
    \includegraphics[width=0.38\linewidth]{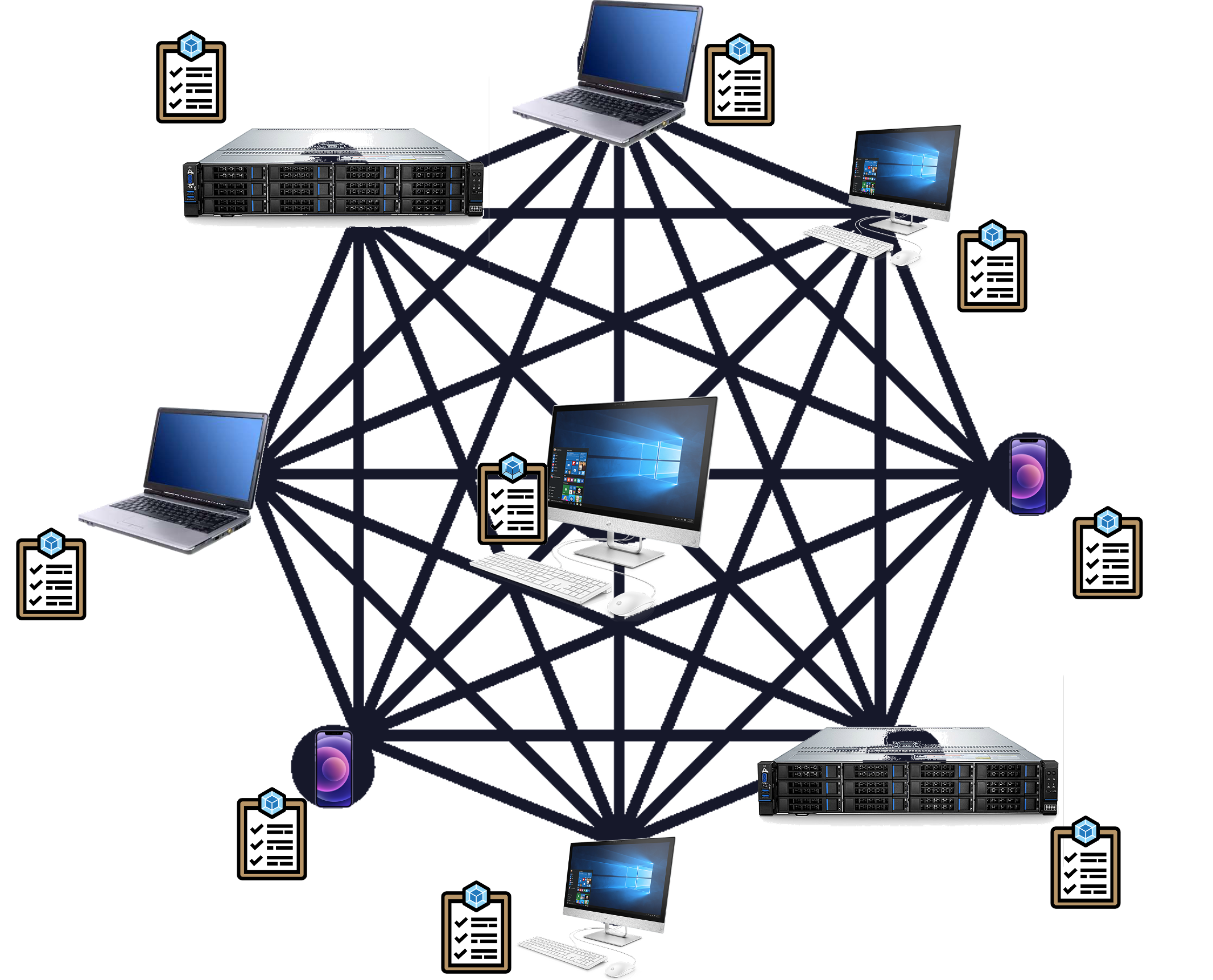}
    \caption{A distributed ledger mirrors each digitized systems' CPS sequence, which in turn may be resolved to a specific Data Proxy mirroring system parameters. }
    \label{dlt}
\end{figure}

The CPS Sequence is associated with the Proxy as part of a universally-unique digital identifier. This identifier may then be used to associate an instance's Proxy with an entry on a Distributed Ledger, similarly to how barcodes or RFID tags are serialized as part of global standards. A representative distributed ledger is shown in Fig.~\ref{dlt}; each node contains a copy of the asset manifest reflecting mirrored systems' CPS sequences and information on where to find and access the associated Data Proxy. 

An overlying architecture, such as Twinbase~\cite{9568895} may be adapted to provide a ``wrapper'' for Data Proxies through which DLT manifest entries may be resolved. The DLT itself may be exposed to an external API depending on application needs. In this manner, Proxies may be searched (with appropriate permission), analyzed, and engaged with at scale.

\begin{figure*}
    \centering
    \includegraphics[width=0.8\textwidth]{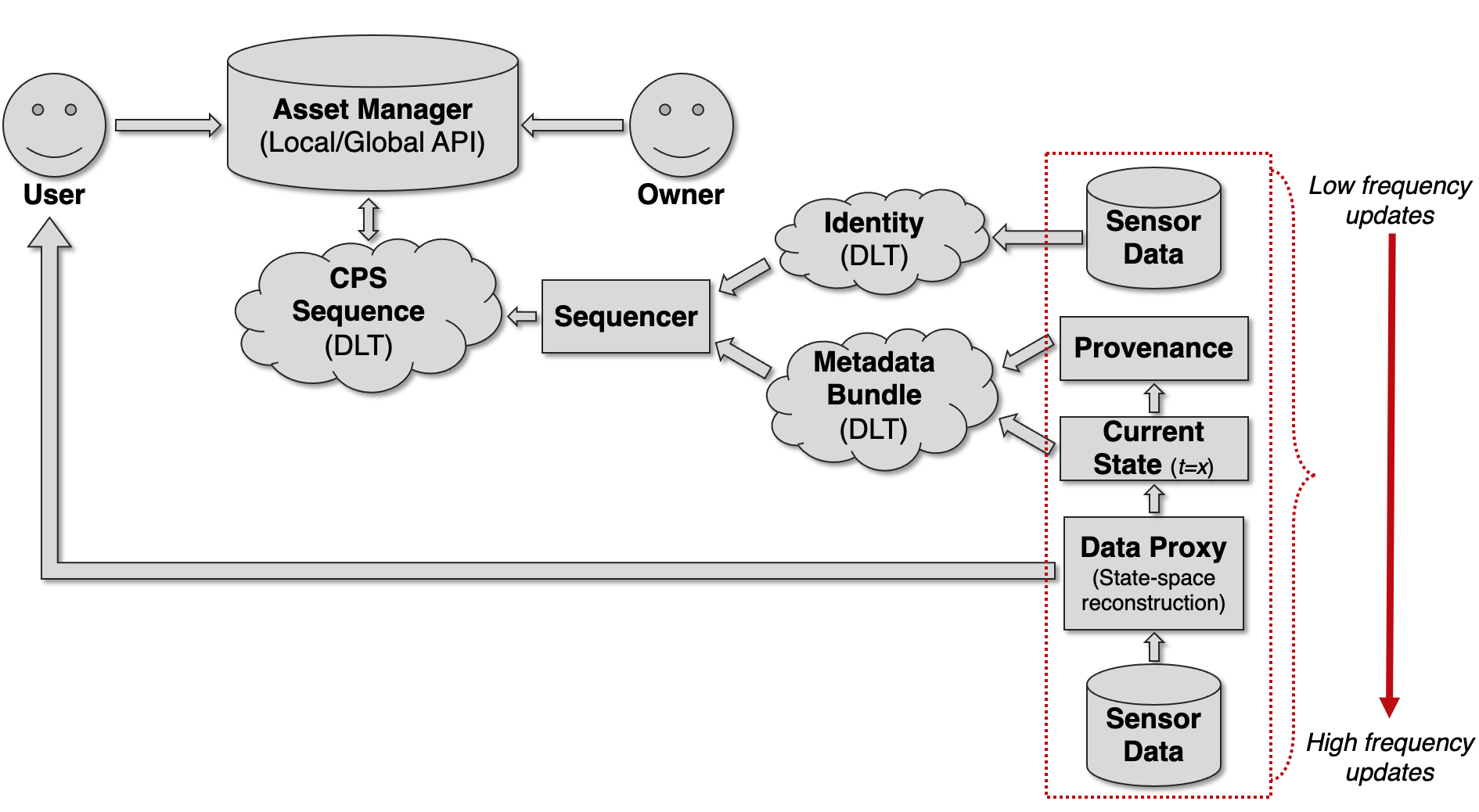}
    \caption{A representative interaction among asset owners and users of asset data. An asset manager manages ownership and permissions, while mediating data access requests for identity, metadata, and Proxy-held information.}
    \label{twinbase_2}
\end{figure*}

A representative embodiment of a Proxy management scheme is shown in Fig.~\ref{twinbase_2}. Here, an asset's Owner may interface with an Asset Manager to assign permissions to assign permissions for Users to engage with their systems. The Asset Manager is a high-level API that ``wraps'' the various DLT's involved in representing CPS Sequences and associated Data Proxy mirrors, and may be public or private in nature and global or local in scale. 

When a User wishes to interact with an asset's metadata or Proxy, the User makes a query with the Asset Manager, which then establishes a connection with the asset's CPS Sequence as stored within a DLT. The CPS Sequence is updated intermittently and reflects the system's Identity, as determined from sensor data at time of origin and stored within a DLT, and the system's Metadata, as determined by its provenance and current state (informed by a Proxy, which itself is a reflection of high-frequency sensor data). 

When the User seeks to monitor the asset's identity or provenance for supply chain and other lower-frequency applications, the Asset Manager shares with the user the CPS Sequence itself. When the User wishes to engage with higher-frequency data, such as might be used in applications typically served by Digital Twins, the Asset Manager shares with the user a connection directly to the Proxy model itself. This is the same model used to estimate the system's current state and to update the Provenance when certain conditions are met. This maintains resource lightweightness and security. 

% User can pull current data; metadata updated less frequently

In our case, the owner of multiple properties might store Proxies of each tenant's keys, and provide tenants access to the Asset Manager to track and manage the transfer of those keys to future tenants, or to recreate the keys from Proxy models in the event of a lockout. The same approach may be used for systems across scales, such as shipping containers, manufacturing equipment, or even factories -- providing secure and efficient visibility while at the same time facilitating safe and secure access to richer system data to enable analytics and other applications.

% Proxy registry may not be globally unique
% Asset manager is the API - easy to use
% DLT stores/returns metadata, CPS sequence

%The system will particularly build upon the concept of hyper-sensitive microclassification and characterization detect miniscule faults and imperfections suitable for unique instance identification,~\cite{siegel2020surveying} akin to the ``witness'' marks lock tumblers may leave on keys, or that certain manufacturing techniques leave on components as a result of process artifacts. Automated system identification may therefore be used to uniquely serialize diverse physical systems without human intervention. Once uniquely and automatically identified, systems are mirrored as Proxies using few computational resources. Low-touch physical system mirroring will support automated asset tracking and supply chain management. With the thoughtful addition of provenance metadata, the concept of DLT and NFTs is ideally suited to storing and serving Data Proxies tied to individual system instance identifiers.

This lightweight and scalable digital asset identification, mirroring, and tracking approach will provide enhanced insight into diverse constrained systems that are otherwise unable to self-report. Growing the mass-scale adoption of mass-scale digital mirroring has the potential to advance the transparency and accountability of diverse supply chains, while simultaneously addressing critical supply chain integrity threats. 

%% file: challenges.tex
The proposed system's scalable approach is suitable for validating authenticity of diverse systems, even those with constrained compute and/or limited to no proprioceptive sensing capabilities. and also creating reasonably accurate digital mirrors suitable for analytics and other applications. This large-scale, lightweight and automated mirroring will improve access to trusted CPS and their benefits, bringing the advantages of Digital Twins (integrity, accountability, authenticity, transparency and insight) to technologically underserved industries. Further, storing unique identifiers in a distributed ledger will allow increased transparency into system history, ownership, and other factors, thereby enabling advanced supply chain and logistics optimization. Such a system provides social benefit by facilitating lifecycle cradle-to-grave environmental monitoring of non-sensored goods and enables the capture of newly-available data, supporting the creation of jobs for the workforce of the future. 

However, there are challenges porting this approach to other physical assets at scale. First, there are environmental consequences to hashing blocks onto the Ethereum network: the energy required for a single transaction is more than a US household consumes in a day\cite{fairley2019ethereum}. Such consumption makes universal asset ``tagging'' infeasible. Another challenge is the increasing cost of listing an asset on the Ethereum network: to incentivize Ethereum miners to hash a block onto the Ethereum chain, a ``gas'' price is paid by the party seeking the transaction's verification\cite{pierro2019influence}. Should everyday objects be digitized, it is unlikely anyone would pay a today-typical \$3.00 fee to have their pencil hashed to the blockchain. 

DLT like directed acyclic graphs (DAGs) offer improved scalability. DAGs do not require hashing fees and need less processing power and therefore energy than conventional approaches. DAGs offer similar benefits of distributed, immutable consensus across a ledger~\cite{kondrudirected} and can be engaged for real-time data transfer and logging~\cite{liu2011supporting}. However, DAGs have fewer established use cases and DAG NFT's must be developed. Other challenges to be addressed include developing a means to replace a ``lost'' token in a cryptographically secure way, and DAG's relatively-lower concensus security (33\% attacks result in compromise). 

Irrespective of the DLT chosen, the proposed approach values security over privacy. Sensitive or private assets are not suitable for storage on a public distributed ledger; in these cases, privacy may be traded for security in the form of private blockchain instances. There is also a need to codify and standardize the reporting requirements for system-impacting events such that metadata might be accurately and comprehensively reflected for diverse systems. 

Finally, the notion of identity must be better understood to create a system that manages unique and diverse assets effectively - that is, if elements of a component or system are replaced, is that system the same? These and other philosophical debates will be part and parcel with the development of any large-scale proof-of-concept for automated distributed asset management. 

%% file: conclusion.tex
Diverse, distributed supply chains invite risk for counterfeit, compromise, or non-compliance. Consistent validation of the trusted identity and integrity of CPS in support of supervision, management, tracking and transferrability mitigates this risk, improving trust in the complex systems comprising national critical functions. Attaching to this identity measures of current system state and provenance enable richer suites of asset management and supervision applications, as well as the development of system analytics and optimization today best suited for Digital Twinning approaches. 

We explored emerging Deep Technology affordances~\cite{siegel2020cultivating} that might facilitate large-scale asset identification and supervision, and building upon these, proposed the development of a system for automatically validating the identity and provenance of diverse, distributed systems. This system comprises distributed system identification and asset integrity tracking techniques built upon fundamental mechanical and software engineering capabilities to create and identify differentiating features for systems with limited internal sensing capabilities. We propose ``CPS sequencing'' as a lightweight and scalable approach for CPS instance identification and automated metadata matching suitable for validating authenticity of diverse systems at scale, even those with constrained compute and/or sensing capabilities. Storing unique identifiers in a distributed ledger will allow increased transparency into system history, ownership, and other factors. These affordances, along with the approach's low resource requirements, will allow for the mass-scale adoption of digital mirroring solutions ill suited for conventional approaches - enabling pervasive identification (similar to RFID, QR codes) with the benefits of richer metadata and realtime insights (similar to those enabled by Digital Twins). 

Implemented well, the proposed approach has the potential to facilitate automated system identification and the ability to pair an asset with a system-specific, immutable digital identifier. Identification and model-matching techniques will form a control loop validating an instance's authenticity and therefore provenance. This form of automated non-repudiable track-and-trace enables trust for CPS' use in critical supply chain and logistics applications related to healthcare, space systems and other national critical functions.

Storing unique identifiers with DLT will allow increased transparency and insight into system history, ownership, and other factors that could be critical to establishing a traceable audit trail, thereby enabling governance procedures\cite{falco2021governing}. Such an approach may be readily adopted by environmental transparency efforts such as cradle-to-grave lifecycle tracking of non-sensored goods; moreover, it will provide logistics and supply chain information where previously there was none, leading to cross-domain insights that may enhance operational efficiency, safety, and security.

As next steps, the authors propose implementing a representative system comprising automated system identification, pervasive mirroring and system digitization, and mirror management with metadata. In parallel, the authors aim to address remaining challenges in architecture design related to defining identify, representing metadata, and improving system generalizability. This will require the development of ultra-low-resource sequencing implementations - perhaps even those with integral ``mining'' components such that asset identification and distributed ledger node service may run on constrained edge devices directly, with sequencing serving as a form of proof-of-work. In this manner, classification, the distributed ledger, and cryptographic processes may be run entirely on the subset of instrumented devices with onboard computation, leading to scalability and resilience benefits. To further enhance the solution's applicability in sensitive domains, the authors will also evaluate operation on proprietary distributed ledgers. This will allow CPS identities and transactions to be made visible only to select entities holding an appropriate digital key.  